# Why are Developers Struggling to Put GDPR into Practice when Developing Privacy-Preserving Software Systems?

*Abdulrahman Alhazmi, Department of Computer Science and IT, School of Engineering and Mathematical Sciences at La Trobe University.*
*Nalin A. G. Arachchilage, Optus La Trobe Cyber Security Research Hub, Department of Computer Science and IT, School of Engineering and Mathematical Sciences.*

## Abstract

The use of software applications is inevitable as they provide different services to users. The software applications collect, store users' data, and sometimes share with the third party, even without the user consent. One can argue that software developers do not implement privacy into the software applications they develop or take GDPR (General Data Protection Law) law into account. Failing to do this, may lead to software applications that open up privacy breaches (e.g. data breach). The GDPR law provides a set of guidelines for developers and organizations on how to protect user data when they are interacting with software applications. Previous research has attempted to investigate what hinders developers from embedding privacy into software systems. However, there has been no detailed investigation on why they cannot develop privacy-preserving systems taking GDPR into consideration, which is imperative to develop software applications that preserve privacy. Therefore, this paper investigates the issues that hinder software developers from implementing software applications taking GDPR law on-board. Our study findings revealed that developers are not familiar with GDPR principles. Even some of them are, they lack knowledge of the GDPR principles and their techniques to use when developing privacy-preserving software systems.

## 1. Introduction

The increased use of software systems has led to users sharing their data widely [1]. The data is shared and accessed in ways users find it difficult to understand [2]. One of the challenges to this can be software applications are not embedded taking privacy into consideration. The GDPR law explains in six elements [5] that provide guidelines to developers enabling for them to develop software systems that preserve privacy. These elements begin with Lawfulness, Fairness, and Transparency, stating collected data should be processed lawfully, fairly and in a transparent manner [5]. The second principle is Purpose Limitation. This principle states that user data should not be used in purposes that are unexpected by the users [5]. The third principle is Data Minimization which states that the only data that is required in a given function will be collected [5]. Accuracy, which is the fourth principle in GDPR, that describes the collected data from data subjects should be kept accurately and always up to date [5]. The fifth principle is Storage Limitation which states the data should not be held by the system if it's no longer needed [5]. The last principle is Integrity and Confidentiality. Confidentiality states that data should not be accessed by unauthorized people while Integrity defines that data should not even be changed by unauthorized access [5]. Software applications that are not being designed to preserve user privacy are prone to data breaches. Previous research has revealed that software developers may not adhere to these principles when implementing privacy-preserving software systems [6] [7]. In addition, there have been data breaches on different software systems that have enormous personal data such as Facebook [3], [4], [9], [10], Yahoo [8] and the most recent case is the Zoom [18], which clearly indicate that developers find it difficult to embed privacy into software applications they develop though the GDPR law came into force [5].

Data breaches occurring but software applications continue to fail protecting user privacy through GDPR could be an issue. One could argue that this can be improved by improving software developers' behavior (i.e. implementation practices) when developing privacy-preserving software systems. On the other hand, they may need help because they are neither privacy nor security experts [11]. Previous research has also revealed that there is currently a lack of resources for software developers, such as process, protocols or techniques in place, to implement privacy into software systems [4] [9] [10] [11]. This brings up a need to investigate why developers cannot embed privacy taking GDPR on board. This paper will investigate developers on the challenges they face when embedding privacy taking GDPR into consideration. Our preliminary study findings revealed that developers are not familiar with GDPR elements. Those who are vaguely familiar with the elements still lack enough knowledge of the elements and their techniques to use when developing privacy-preserving software systems.

## 2. Literature Review

In the past, researchers have attempted to investigate some of the General Data Protection Regulation (GDPR) principles. However, no research has been done to investigate all of the principles. Van et al. [16] have investigated developers' attitudes towards handling personal data considering its lawfulness, Fairness, and Transparency (i.e. first principle of GDPR) as well as how developers' use data minimization technique (i.e. second principle of GDPR) when developing privacy-preserving software systems. They developed a scale to measure developer attitude for handling the personal data of the user. In their study they focused on three main points, informed consent, data minimization, and data monetization. They found that there exist mismatches between developer's attitudes and their self-perceived behaviors. However, authors failed to consider other principles covered in GDPR such as is Purpose Limitation, Accuracy, Storage Limitation and Integrity and Confidentiality. As stated, which are imperative to investigate when software developers asked to embed privacy into the systems they develop. One can argue that to preserve user privacy in software applications, GDPR law must be complied with[5].

Senarath and Arachchilage also conducted an experiment using 36 software developers in a software design task with instructions to embed privacy in order to identify the problems they face [4]. Their investigation focused only on data minimization principle of the GDPR [4]. Schwerin [5] claims that investigating only data minimization technique, doesn't address the problem from a broader perspective. In other words, failure to implement one GDPR element, is failure to comply with the GDPR law. Research has been carried out to understand how and what affects developers in developing a privacy-preserving software system [19]. The research [12] was more concerned with organization privacy practices. It gives in-depth environmental mechanisms and identifies the environmental components that impact and, are affected by developers when dealing with privacy concerns. It further identifies organizational privacy climate as a good tool that organizations can use to control developers with regard to specific translations of privacy. (Sheth et al, 2014) [13] studied how user and developer perceive privacy. They used a survey study to find out the user response regarding their concern for privacy. Their surveys showed that most of the users are concerned with their personal data and location rather than the interaction data. They also found that people from different regions have variations in their concerns regarding privacy. On the other hand, they found that developers most of the time focused on anonymization and technical measures. (Ayalon, et al, 2017) [14] conducted an online survey to find professional privacy attitudes and practices of developers. They found that developers privacy decision is comprised of different factor which includes organizational privacy, professional and personal perceived privacy. Similarly, the researchers did not attempt to investigate the challenges which are faced by developers when implementing privacy and taking GDPR law on board.

As evidenced by the review above, previous researchers have investigated the issues faced by developers when implementing privacy in software applications. Some of them have attempted to investigate GDPR, however, not in full [4] [16]. To preserve user privacy when they are interacting with software applications, all the principles of GDPR must be implemented [5]. This research empirically investigates the issues software developers face when they attempt to embed privacy into software systems complying with the six elements of GDPR law. This current study will focus on software developers with industry experience in end-user software application development.

Previous researchers have investigated the issues faced by developers when implementing privacy in software applications. Most of them have partially attempted to investigate GDPR law. However, there is the need for a more comprehensive study that investigates all the principles.

Previous studies clearly show that there are differences in the user and developers' perception regarding software privacy. These differences can be seen in surveys carried out by researchers. To overcome these differences in the developer and user environment, all the principles of GDPR must be implemented.

## 3. Methodology

To investigate why developers are unable to use GDPR when implementing privacy in software systems, we conducted interviews with six developers who were identified on LinkedIn. The participants (four males and two females) were all experienced in software development. Three participants were from Europe while the remaining two from Australia and one from Africa. Before the interviews commenced, the project was approved by the Latrobe University Ethical Board. We sent a consent form to each participant, to have their permission to include them in the research. The interviews were conducted remotely via Zoom. We sent a scenario and UML diagrams to the participants which were developed with the reference to GDPR. The interview questions were about the scenario and the diagrams.

The participants were asked questions that enabled us to investigate how they implement each GDPR principle. We asked the participants if they knew any data privacy law.

This was to give us an insight into whether they were familiar with the GDPR. The participants were asked different sets of questions, aimed at investigating different principle of GDPR. For instance, whether they can inform data subjects of a data breach and how long they would take to inform them. This was aimed at investigating the first GDPR principle i.e. Lawfulness, Fairness, and Transparency.

The Data minimization principle was investigated by asking the participants which set of data they will collect from users in the scenario given. Our questions to the participants enabled us to investigate the six principles of GDPR. We asked "why" follow-up questions from the answers we got from the participants to get more information related to their understanding of GDPR e.g. when the answer was 'I can collect name, username, and password', we asked 'why collect this'? We wrote down the responses of the developers as the interview was ongoing. We applied semantic analysis when writing the responses in order to write sentences that clearly point out what participants meant in their speech. Semantic analysis refers to the use of contextual clues surrounding the words and phrases of a text to better understand the practical meaning of the content of that text [15].

We applied grounded theory [17] to analyze the participants' descriptive answers. Some of the answers were 'yes/no' and did not require grounded theory to analyze. Responses from all the participants were analyzed. Three coding schemes were applied to the data collected [17]. This was important as it enabled us to capture all the issues that developers face. The first coding we applied is open coding [17]. In this coding scheme, we read through the answers several times created tentative labels for chunks of data to summarize the responses. The second coding scheme, axial coding, was applied to identify the relationship between the open codes [17]. We also categorized similar answers in this stage and were able to come up with many categories. To link these categories into fewer categories that enabled us to analyze the data more accurately, we used the third coding which is selective coding [17]. From this analysis, we were able to identify why developers are unable to use GDPR when embedding privacy into software systems

### 4. Results and Discussion

The study findings revealed two major issues that prevent developers from embedding privacy using GDPR law. The issues are discussed below.

4.1 **Lack of having good techniques to implement GDPR law.** Our study findings revealed that four out of six participants lacked the techniques to implement GDPR principles effectively. Storage Limitation and Purpose Limitation were the leading principles for which the participants lacked implementing techniques. For example, developer #3 said "I don't have any idea on how to implement storage limitation and purpose limitation. I even do not know what that means". He said that the terms were new to him. Two out of the four had additional challenges in implementing Lawfulness Fairness and Transparency and Accuracy principles.

Moreover, we found out that participants could not differentiate between different techniques required to implement the different GDPR principles. For example, they failed to identify the techniques and elaborate on how they differ from implementing "Purpose Limitation" to Lawfulness, Fairness, And Transparency, according to GDPR.

Three participants identified different techniques on Lawfulness, Fairness, And Transparency principle but when asked questions about Purpose Limitations, they responded that they had already answered that question when responding to the question on the Lawfulness, Fairness, And Transparency. For example, developer #5 said "I don't have to answer how I can ensure purpose limitation as I already responded to this on lawfulness, fairness, and transparency". All six participants had techniques such as encryption and encapsulation to implement the integrity and Confidentiality principle. They also talked about techniques to implement the Data Minimizations principle as follows:

(a) Developers #1 and #6: Collection of data once when needed

(b) Developer #2: Analysis of the process to help identify all the requirements of the system

(c) Developer #3: Collection of critical data after analyzing the process

(d) Developer #4: Analyses of the functional process

(e) Developer #5: Analysis of the system requirements to come up with the data.

The representations given below indicates the number of participants that had challenges in different GDPR elements.

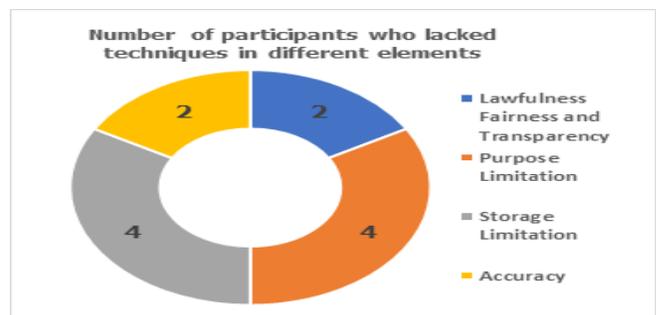

*Figure 1: Number of participants that had challenges in implementing the different GDPR principles*

4.2 **Participants were not familiar with GDPR law and lacked proper guidelines to implement it.** Our study revealed that five out of six participants never had an idea of the GDPR law. Developers from Europe had greater understanding of GDPR elements. Some knew different data privacy laws e.g. one developer said that his country has some data privacy laws which he uses when developing software applications. From the explanation given by the developer, his country's privacy laws were more concerned with integrity and confidentiality and therefore, doesn't address the GDPR law in full. Our study revealed that this resulted in them having issues with the other five GDPR such as is Lawfulness, Fairness and Transparency, Data Minimization, Purpose Limitation, Accuracy, and Storage Limitation.

Participants were also unable to implement the Purpose Limitation principle because they were not familiar with it. For example, developer #5 said he was hearing about this principle for the first time, and therefore, did not know how to address implement it.

Our study revealed that the GDPR law is not well known to the software developers and those familiar with it, did not understand all the principles. For example, one developer who was familiar with the GDPR law said that he only heard about the Integrity and Confidentiality and Data Minimization principles. Our study found this to be the main problem as none can implement something that s/he is not familiar with. The representation below indicates the familiarity of participants with data laws. Only one participant was familiar with the GDPR law. Two participants were familiar with some other data law such as their country's data law. However, three participants stated that they were not familiar with any data laws.

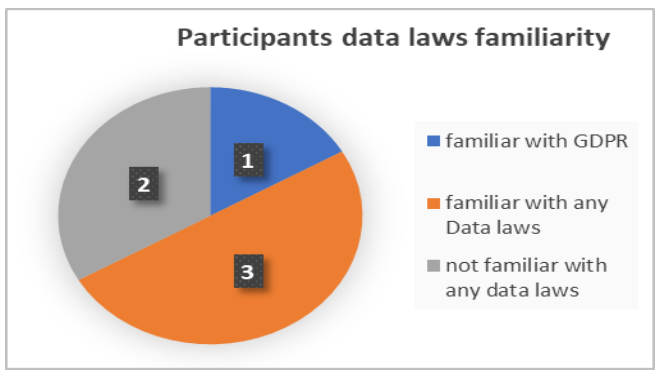

*Figure 2: Participants' familiarity with data laws*

## 5. Recommendations

We recommend the following guidelines to address the issues identified in our investigation:

- The universal standard of GDPR law should be made available, so software creators can follow (irrespective to the country they belong or the software application they develop for) while maintaining consistency; all nations should adopt and enforce it to enable developers to familiarize themselves with the law and be bound by it.
- GDPR law should be accompanied by guiding techniques for each principle to ensure effective implementation of the principles and to prevent developers from using sub-standard techniques. Developers should be given formal knowledge of the GDPR law and processes and techniques to apply them when developing privacy-preserving software systems.

These recommendations will help to address the issues that we identified in our study. Our study contributes to improving the privacy of personal data taking GDPR on board. If developers comply with GDPR law, data breaches may be reduced.

## 6. Conclusion and Future Work

This research sought to identify the issues relating to the inability of software developers to embed privacy into software, taking GDPR into consideration. The findings of this research enabled us to derive recommendations that would effectively support developers when they embed privacy to take GDPR law on board.

This research revealed the following about most of the developers:
A. They are not familiar with GDPR principles
B. They lack formal knowledge of the six principles
C. They have no knowledge or idea of how to apply the GDPR principles into practice due to limited resources (i.e. tools, techniques, or processes to put GDPR into practice).

Future studies on the current topic are therefore recommended. We are planning to conduct qualitative analysis of this study with more participants, investigating why software developers cannot develop software applications that preserve user privacy. In addition, we seek to identify tools, techniques and processes that are required for them to put GDPR into practice.

# 7. References

[1] S. Pearson and A. Benameur, "Privacy, security and trust issues arising from cloud computing," in 2010 IEEE Second International Conference on Cloud Computing Technology and Science, pp. 693–702, IEEE, 2010.

[2] M. Van Kleek, I. Liccardi, R. Binns, J. Zhao, D. J. Weitzner, and N. Shadbolt, "Better the devil you know: Exposing the data sharing practices of smartphone apps," in Proceedings of the 2017 CHI Conference on Human Factors in Computing Systems, pp. 5208–5220, 2017.

[3] C. Cadwalladr and E. Graham-Harrison, "Revealed: 50 million facebook profiles harvested for cambridge analytica in major data breach," The guardian, vol. 17, p. 22, 2018.

[4] A. Senarath and N. A. Arachchilage, "Why developers cannot embed privacy into software systems? an empirical investigation," in Proceedings of the 22nd International Conference on Evaluation and Assessment in Software Engineering 2018, pp. 211–216, 2018.

[5] S. Schwerin, "Blockchain and privacy protection in the case of the european general data protection regulation (gdpr): a delphi study," The Journal of the British Blockchain Association, vol. 1, no. 1, p. 3554, 2018.

[6] Li ZS, Werner C, Ernst N, Damian D. GDPR Compliance in the Context of Continuous Integration. arXiv preprint arXiv:2002.06830. 2020 Feb 17.

[7] S. Wachter, "Gdpr and the internet of things: Guidelines to protect users' identity and privacy," 2018.

[8] L. J. Trautman and P. C. Ormerod, "Corporate directors' and officers' cybersecurity standard of care: The yahoo data breach," Am. UL Rev., vol. 66, p. 1231, 2016.

[9] A. Senarath and N. A. G. Arachchilage, "A data minimization model for embedding privacy into software systems," Computers & Security, vol. 87, p. 101605, 2019.

[10] A. Senarath and N. A. G. Arachchilage, "Understanding software developers' approach towards implementing data minimization," USENIX Symposium on Usable Privacy and Security (SOUPS) 2018. August 12–14, 2018, Baltimore, MD, USA. arXiv preprint arXiv:1808.01479.

[11] C. Wijayarathna, M. Grobler, and N. A. Arachchilage, "Software developers need help too! developing a methodology to analyse cognitive dimension-based feedback on usability," Behaviour & Information Technology, pp. 1–22, 2019.

[12] I. Hadar, T. Hasson, O. Ayalon, E. Toch, M. Birnhack, S. Sherman, and A. Balissa, "Privacy by designers: software developers' privacy mindset," Empirical Software Engineering, vol. 23, no. 1, pp. 259–289, 2018.

[13] S. Sheth, G. Kaiser, and W. Maalej, "Us and them: a study of privacy requirements across north america, asia, and europe," in Proceedings of the 36th International Conference on Software Engineering, pp. 859–870, 2014.

[14] O. Ayalon, E. Toch, I. Hadar, and M. Birnhack, "How developers make design decisions about users' privacy: The place of professional communities and organizational climate," in Companion of the 2017 ACM Conference on Computer Supported Cooperative Work and Social Computing, pp. 135–138, 2017.

[15] Y. Gong and X. Liu, "Generic text summarization using relevance measure and latent semantic analysis," in Proceedings of the 24th annual international ACM SIGIR conference on research and development in information retrieval, pp. 19–25, 2001.

[16] van der Linden, D., Hadar, I., Edwards, M., & Rashid, A. "Data, data, everywhere: quantifying software developers' privacy attitudes," 09 2019.

[17] Charmaz, K., & Belgrave, L. L. (2007). Grounded theory. The Blackwell encyclopedia of sociology.

[18] Khan, N. A., Brohi, S. N., & Zaman, N. (2020). Ten Deadly Cyber Security Threats Amid COVID-19 Pandemic.

[19] Senarath, A., Grobler, M., & Arachchilage, N. A. G. (2019). Will they use it or not? Investigating software developers' intention to follow privacy engineering methodologies. ACM Transactions on Privacy and Security (TOPS), 22(4), 1-30.